\documentclass[multphys,vecphys]{svmult}

\newcommand{\OI}{O\,{\sc i}}
\newcommand{\OIII}{O\,{\sc iii}}
\newcommand{\CII}{C\,{\sc ii}}

\newcommand{\NII}{N\,{\sc ii}}
\newcommand{\NIII}{N\,{\sc iii}}

\usepackage{makeidx}     
\usepackage{graphicx}    
\usepackage{multicol}    

\makeindex             

\begin{document}

\title*{Far--IR spectroscopy towards Sagittarius B2}
\author{Javier R. Goicoechea \inst{1}\and Jos\'e Cernicharo\inst{2}}
\institute{Departamento de Astrof\'{\i}sica Molecular e Infrarroja,
   IEM/CSIC, Serrano 121, 28006 Madrid, Spain.
\texttt{javier@damir.iem.csic.es}
\and Departamento de Astrof\'{\i}sica Molecular e Infrarroja,
   IEM/CSIC, Serrano 121, 28006 Madrid, Spain. 
   \texttt{cerni@damir.iem.csic.es}}
%
%
\maketitle

The far--IR is a unique wavelength range for Astrophysical studies,
however, it can only be fully sampled from space platforms.
The fundamental rotational transitions of light molecules,
the high--$J$ transitions of polyatomic species, the bending modes
of non-polar molecules, several atomic fine structure lines
and many frequencies blocked by the earth atmosphere can
only be observed between 50 and 200~$\mu$m (6.0 and 1.5~THz).
In this contribution we present the far--IR spectrum of Sgr~B2
at a resolution of $\sim$35~km~s$^{-1}$, the 
\textit{``Rosetta stone''} of ISO's spectra.
We also discuss the perspectives of the far--IR Astronomy 
in the context of the future telescopes under development.

\section{Introduction}
\label{sec:1}

Due to the atmospheric opacity, the far--IR domain has been the last window
used in Astronomy. 
For the first time, the \textit{Infrared Space Observatory} (ISO) has 
opened this spectral frequency range  
through \textit{molecular spectroscopy}. 
The sensitivity of the instrumentation on board 
this platform  has no comparison with the few previous space missions
or airborne observations carried before the launch of ISO.
Almost all the operative range of ISO in the far--IR
was not explored before.
The far--IR spectrum of the most significant galactic sources was
unknown. In particular, the main radiation emitters, the molecules, 
were unidentified. Far--IR observations are specially suitable to the 
study of the \textit{warm gas} in molecular clouds.
Among these sources, Sgr~B2 in the
Galactic Center, is a  paradigmatic object for our knowledge of
the \textit{chemical complexity} of the Galaxy.

The molecular species 
and the atomic fine structure lines that can be detected in the 
far--IR domain are essential for the study 
of the physical and chemical conditions of the interstellar medium. 
The bulk of these species can only be observed
from space platforms.
As a example, the water vapor abundance can determine if stars will be 
formed during the gravitational collapse of a molecular cloud.
Another example are the non--polar  carbon chains. These species can be the
\textit{``skeletons''} from which the large carbon molecules
responsible of a great part of the  IR emission in the Galaxy 
can be formed. Due to the lack of permanent electric dipole, these
species do not have rotational spectrum to be observed from radio
telescopes.

\clearpage

In this contribution we present the main results of our detailed study of
the \textit{Long--wavelength spectrometer} (LWS)  spectrum 
of Sgr~B2(M) between 43~$\mu$m (7.0~THz) and 197~$\mu$m (1.5~THz). 
Both with the grating ($\lambda/\Delta\lambda$$\sim$200) and
with the  Fabry--Perot (FP; $\lambda/\Delta\lambda$$\sim$10000).

\section{The far--IR spectrum of Sgr B2(M)}

\begin{figure}
\centering
\includegraphics[height=10.5 cm, angle=-90]{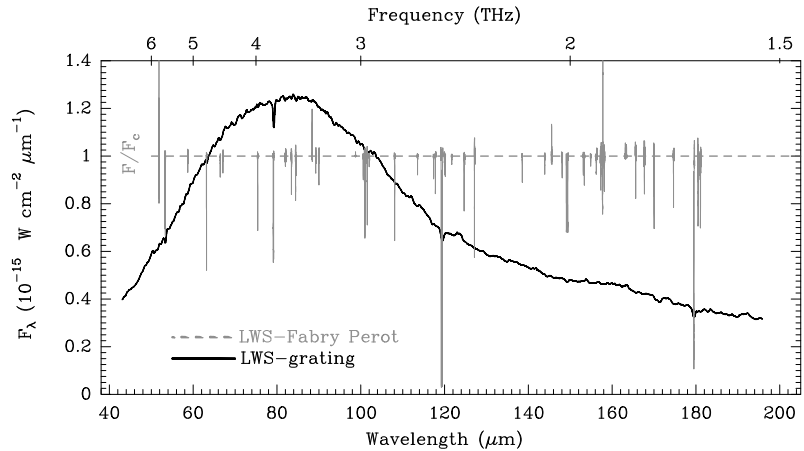}
\caption{LWS/grating spectrum of Sgr~B2(M) 
($\alpha$=17$^h$44$^m$10.61$^s$, $\delta$=-28$^o$22$^{'}$30.0$^{''}$ [J1950])
between  43 and  197~$\mu$m and main detections with the LWS/FP.
Ordinate is valid for the  grating (in 
W~cm$^{-2}$~$\mu$m$^{-1}$) and for the  FP observations (in  F/F$_c$).}
\label{fig:1}
\end{figure}

Figure \ref{fig:1} shows the grating and FP observations of Sgr~B2(M).
The far--IR spectrum of Sgr~B2 is dominated by the high thermal emission
of the dust ($\sim$28000 Jy at $\sim$80~$\mu$m [$\sim$3.75 THz]) and
by the molecular/atomic  lines.

\begin{figure}
\centering
\includegraphics[height=11 cm, angle=90]{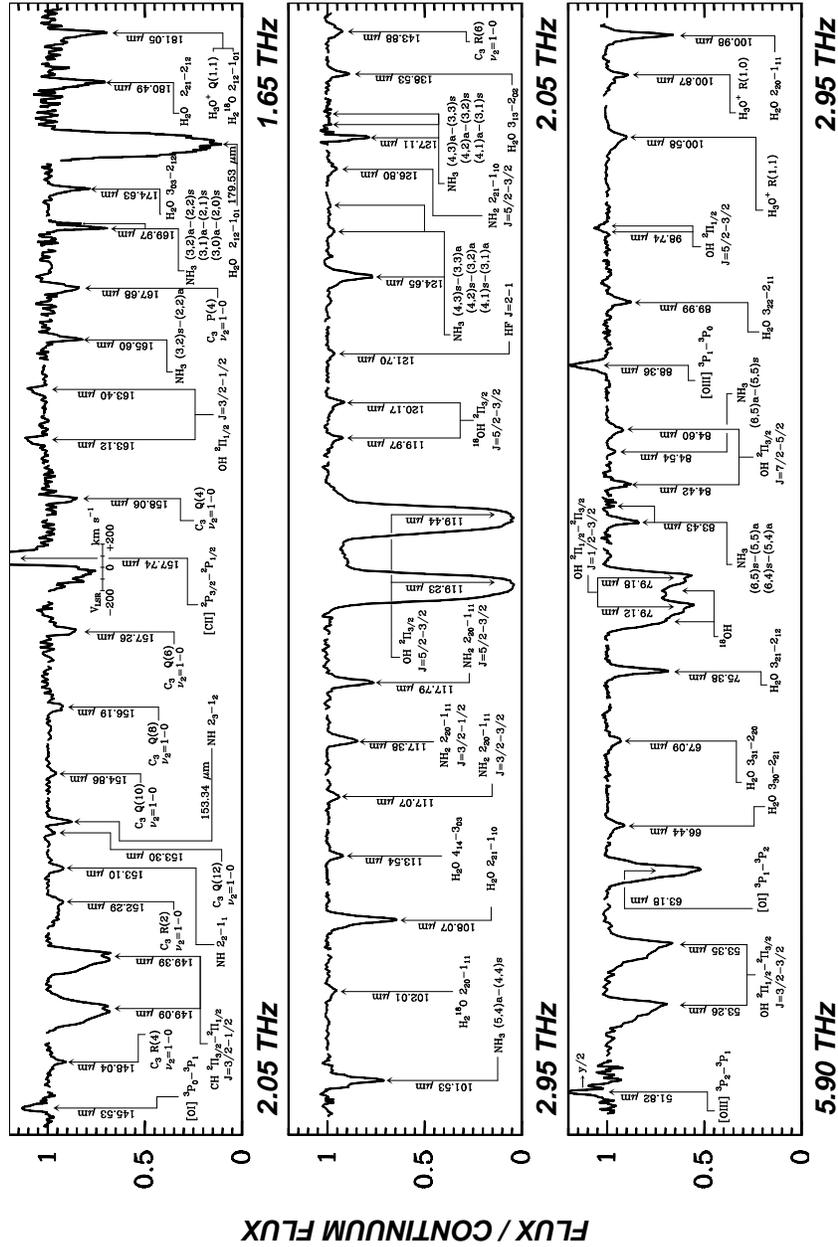}
\caption{Far--IR features in the spectrum of Sgr~B2(M) taken with the
ISO/LWS Fabry-Perot. Note the wavelength discontinuity of the spectrum 
after each line.}
\label{fig:2}        
\end{figure}

Figure \ref{fig:2} shows the most abundant species that can be
detected with ISO in the far--IR (see Goicoechea et al. 2004) 
and gives insights of which could be detected  in other molecular 
clouds of the interstellar medium (ISM).

The molecular richness in the outer layers of Sgr~B2 is  probed by the
FP detections towards Sgr~B2(M),
where more than 70 lines from 15 molecular and
atomic species are observed at high signal to noise ratio.

The spectral lines that  appear in the far--IR and that have
been observed in Sgr~B2 can be classified in:

\begin{itemize}

\item  \textbf{Rotational lines}
of light \textit{O--bearing} molecules such as H$_2$O, H$_2^{18}$O, 
OH, $^{18}$OH, and H$_3$O$^+$, \textit{N--bearing} molecules such as 
NH, NH$_2$ and NH$_3$ and other diatomic species such as CH, HD or HF.
\item  \textbf{Bending modes} of non-polar 
species such as the C$_n$ carbon chains.

\item \textbf{Atomic fine structure lines} of 
[\OI], [\OIII], [\CII], [\NII] and [\NIII].

\end{itemize}

\clearpage

\clearpage

The far--IR observations have offered the opportunity of
studying the chemistry of $O$-- and $N$--\textit{bearing} species
through the  H$_3$O$^+$/H$_2$O/OH/O 
(Goicoechea \& Cernicharo 2001, 2002) and NH$_3$/NH$_2$/NH observations.
In addition, the abundance  of non--polar molecules
such as C$_3$ have been determined for the
first time  (Cernicharo et al. 2000). Finally, the influence 
of the UV radiation field in  the external layers of Sgr~B2 
(the environment of the cloud) has been shown through the analysis of
the atomic fine structure lines (Goicoechea et al 2004).
All these new scientific goals can be obtained only  through space
observations in the far--IR spectral range.

\section{Future Perspectives}

The spatial scales of the physical and kinetical phenomena within
molecular clouds and the amount of molecular species that
could be present, claim for a much better sensitivity and
spectral/spatial resolution in the far--IR.
In the next years, most of these handicaps will be
overcome by \textit{ALMA} ($\sim$2010) for interferometric observations
with $\lambda>350$~$\mu$m. 
On the other hand,  the  \textit{Herschel Space Telescope} ($\sim$2007)
will offer much better spectral resolution than ISO
for $\lambda>157$~$\mu$m.
However, the spatial resolution will be similar to that
of current millimeter single--dish telescopes.

In any case, the full spectral window between 50 and 200~$\mu$m 
(6.0 and 1.5~THz) is not going to be  sampled by any planned space mission 
in the next decade. Taking into account all the background
learned from the ISO/LWS observations (which is
only the top of the iceberg), the request of high spectral and spatial
resolution is mandatory in order to understand and get deeper into
the unique phenomena occurring in the far--IR domain. The high spectral
resolution ($<1$~km~s$^{-1}$) can
be achieved by a single-dish heterodyne instrument, while the
high spatial resolution can only be achieved with
a large single dish telescope with  direct/heterodyne detection 
techniques or with a heterodyne space interferometer.

\paragraph{Acknowledgements}

  We thank Spanish DGES and PNIE for funding
support under grant PANAYA 2000--1784 and ESP2001-4516.
We also thank Nemesio J. Rodr\'{\i}guez-Fern\'andez 
for stimulating discussions about Sgr B2 and the Galactic Center.

%
%
 \bibliographystyle{}
 \bibliography{}
%


\printindex
\end{document}